# Properties of Quasi-Oscillator in Position-Dependent Mass Formalism

S. Zare[1*] and H. Hassanabadi[1]

[1]*Physics Department, Shahrood University of Technology, Shahrood, Iran*

*[soroushzrg@gmail.com](mailto:soroushzrg@gmail.com)

**Abstract**

After introducing Schrödinger equation within position- dependent mass formalism, a quasi-oscillator has been considered. Eigen functions and energy spectra have been obtained analytical. Then thermodynamic properties, information entropy and uncertainty in coordinate and momentum corresponding the considered system have calculated as well as some depicted.

**Keywords**: Quasi-Oscillator, Thermodynamic properties, position-dependent, uncertainty,

## Introduction

Position-dependent mass property has a wide range of applications in various areas of material science and condensed matter physics [1-8] that's why that many interests of physics have been attracted to this topic [9-15]. As one the most important application of position-dependent mass formalism, we can mention in micro fabrication techniques such as molecular-beam epitaxy and nanolithography [16-18]. Schrödinger equation considering this formalism has been investigated via may different approaches such as path integral [19], super symmetric quantum mechanics [20], Darboux transformation [21], the de Broglie–Bohm approach [22] and Hamiltonian factorization [23]. As a few indications to scientific papers in this literature: Guo-Hua et al [24], have presented the Shannon entropy for the position and momentum eigenstates of the position- dependent Schrödinger equation for a particle with a non-uniform solitonic mass density in the case of a hyperbolic-type potential. Falaye et al [25], considered a particle in position-dependent mass formalis with a nonuniform solitonic mass density and a special squared hyperbolic cosecant potential as an interaction. Amir and Iqbal [26], quantized the classical oscillator by considering the symmetric ordering of operator equivalents of momentum and position-dependent mass respectively and obtained a quantum Hamiltonian which was manifestly Hermitian in the configuration space. With inspiration of above points, we are interested in considering Schrödinger equation within position-dependent mass formalism and a quasi-harmonic potential as an interaction. Then we will analytically investigate this system by obtaining wave function and energy spectra, Shannon information entropy, thermodynamics properties. We have organized this article as follows: Introducing position- dependent Schrödinger equation and quasi-oscillator in detail in Sec. 2. Obtaining wave function and energy spectra corresponding the considered system, in Sec. 3. Calculation of thermodynamic properties of the system in Sec. 4. Some discussions around information entropy in Sec. 5 and in the last section some expectation values and uncertainty principle, are evaluated.

## 2- The one-dimensional Schrödinger equation with position-dependent mass

Then one-dimensional Schrödinger equation with position- dependent mass can be written as [27]:

$$-\frac{\hbar^2}{2m(x)}\frac{\partial^2 \Psi(x)}{\partial x^2} - \frac{3\hbar^2}{4}\left[\frac{d}{dx}\left(\frac{1}{m(x)}\right)\right]\frac{\partial \Psi(x)}{\partial x} + \left(V(x) - \frac{\hbar^2}{4}\left[\frac{d^2}{dx^2}\left(\frac{1}{m(x)}\right)\right]\right)\Psi(x) = E\Psi(x) \quad (1)$$

Where the atomic unit $\hbar = 1$ and $m_0 = 1$ are employed. Performing the following transformations on Eq. (1)

$$\Psi(x) = m(x)^{\frac{1}{2}}\phi(y(x)) \quad (2a)$$

$$\frac{dy}{dx} = m(x)^{\frac{1}{2}} \quad (2b)$$

$$m(x) = \frac{m_0}{(1+\gamma x^2)^2} \quad (2c)$$

and by attention pervious equation we know that [28-30]

$$y(x) = \frac{\sqrt{m_0}}{\sqrt{\gamma}} Arctan(x\sqrt{\gamma}) \quad (3)$$

then, we have

$$-\frac{1}{2}\frac{d^2\phi(y)}{dy^2} + V(y)\phi(y) = E\phi(y) \quad (4)$$

We define

$$V(y) = \frac{1}{2}m_0\omega^2 y^2 \quad (5)$$

on the other hand by using Taylor series in equation (3) and substituting into equation (5), above equation becomes

$$V(x) = \frac{1}{2}m_0\omega^2 x^2 - \frac{1}{3}m_0\gamma\omega^2 x^4 + \frac{23}{90}m_0\gamma^2\omega^2 x^6 \ldots \quad (6)$$

when the γ lead to zero, only the first term in above equation remains. Therefore we can write

$$V(x) = \frac{1}{2}m_0\omega^2 x^2 \quad (7)$$

In Fig.-1 we have plotted the potential versos x for some different γ values.

as the quasi-harmonic interaction. Now, we are in a position to take the next step.

## 3-The wave function and Energy for oscillator

We can get the Weber's differential equation using Equation (4). Substituting equation (5) into (4), we will have Weber's differential equation as a following form [31]:

$$\frac{d^2\phi(y)}{dy^2} + (2E - \omega^2 y^2)\phi(y) = 0 \quad (8)$$

Then by getting $s = \omega y^2$ and substituting into above equation can be written as:

$$s\frac{d^2\phi(s)}{ds^2} + \frac{1}{2}\frac{d\phi(s)}{ds} + \left(\frac{E}{2\omega} - \frac{s}{4}\right)\phi = 0 \quad (9)$$

In order to rewrite Equation (9) in the standard form we split off the asymptotic solution and hence try a solution of the form $\phi(s) = e^{-\frac{s}{2}}\chi(s)$ which leads to

$$s\frac{d^2\chi(s)}{ds^2} + \left(\frac{1}{2} - s\right)\frac{d\chi(s)}{ds} + \left(\frac{E}{2\omega} - \frac{1}{4}\right)\chi = 0 \quad (10)$$

We identify the above equation as the Kummer differential equation (10). The eigenfunction may be expressed in terms of regular confluent hypergeometric functions $M(a,c,s)$ as:

$$\chi(s) = A\,M\left(a, \frac{1}{2}, s\right) + Bs^{\frac{1}{2}}M\left(a + \frac{1}{2}, \frac{3}{2}, s\right) \quad (11)$$

Where A and B are arbitrary constants and

$$a = -\left(\frac{E}{2\omega} - \frac{1}{4}\right) \quad (12)$$

In terms of the variable $y$, the wavefunctions can be written as:

$$\phi(y) = A\, e^{-\frac{1}{2}\omega y^2} M\left(a, \frac{1}{2}, \omega y^2\right) + B(\omega y^2)^{\frac{1}{2}} M\left(a + \frac{1}{2}, \frac{3}{2}, \omega y^2\right) \qquad (13)$$

Finally by attention to equation (2a), (2c) and (3) eigenfunction $\Psi(x)$ becomes:

$$\Psi(x) = \sqrt{\frac{m_0}{(1+\gamma x^2)^2}} \left( A\, e^{-\frac{\omega m_0}{2\gamma} Arctan^2(x\sqrt{\gamma})} M\left(a, \frac{1}{2}, \frac{\omega m_0}{\gamma} Arctan^2(x\sqrt{\gamma})\right) \right.$$

$$\left. + B \frac{\sqrt{\omega m_0}}{\sqrt{\gamma}} Arctan(x\sqrt{\gamma}) M\left(a + \frac{1}{2}, \frac{3}{2}, \frac{\omega m_0}{\gamma} Arctan^2(x\sqrt{\gamma})\right) \right) \qquad (14)$$

The confluent Hypergeometric function are related to the Hermit polynomials through the following equations:

$$H_{2n}(\xi) = (-1)^n \frac{(2n)!}{n!} M\left(-n, \frac{1}{2}, \omega y^2\right) \qquad (15)$$

$$H_{2n+1}(\xi) = (-1)^n \frac{2(2n+1)!}{n!} M\left(-n, \frac{3}{2}, \omega y^2\right) \qquad (16)$$

In view of the above equations, the even and odd eigenfunction may be expressed as

$$\psi_{even} = N_n e^{-\frac{\omega y^2}{2}} H_{2n}(y\sqrt{\omega}) \qquad (17a)$$

$$\psi_{odd} = N_n e^{-\frac{\omega y^2}{2}} H_{2n+1}(y\sqrt{\omega}) \qquad (17b)$$

with the normalization constant given by :

$$N_n = \sqrt{\sqrt{\frac{\omega}{\pi}} \frac{1}{2^2 n!}} \qquad (18)$$

However the even odd eigenfunction may be combined and the stationary states of the relativistic oscillator are

$$\Psi_n(x) = N_n \sqrt{\frac{m_0}{(1+\gamma x^2)^2}}\, e^{-\frac{\omega m_0}{2\gamma} Arctan^2(x\sqrt{\gamma})} H_n\left(\frac{\sqrt{\omega m_0}}{\sqrt{\gamma}} Arctan(x\sqrt{\gamma})\right) \qquad (19)$$

The energy eigenvalues of spin zero particles bound in this oscillator potential may be found using equation (11).
Therefore energy for Even and odd states can be written as:

$$E = \omega\left(n + \frac{1}{2}\right) \quad (20)$$

## 4-Thermodynamic properties of system

In order to consider thermodynamic properties of neutral particle, we concentrate, at first, on the calculation of the partition function [32]

$$Q_1 = \sum_{n=0}^{\infty} e^{-\beta E_n} = e^{-\frac{\beta\omega}{2}} + e^{-\frac{3\beta\omega}{2}} + \cdots = \frac{1}{e^{\frac{\beta\omega}{2}} - e^{-\frac{\beta\omega}{2}}} = \left(2\sinh\frac{\beta\omega}{2}\right)^{-1} \quad (21)$$

Where $\beta = 1/KT$ the partition function for N-body system with no interaction inside obtain via

$$Q_N = (Q_1)^N = \left(2\sinh\frac{\beta\omega}{2}\right)^{-N} \quad (22)$$

Where

$$lnQ_N = -N \ln\left(2\sinh\frac{\beta\omega}{2}\right) \quad (23)$$

Once the Helmholtz free energy is obtained, the other statistical quantities are obtained in a straightforward manner as

$$A = -\frac{1}{\beta} lnQ_N = \frac{N}{\beta} \ln\left(2\sinh\frac{\beta\omega}{2}\right) \quad (24)$$

Chemical potential can obtain as

$$\mu = \frac{\partial A}{\partial N} = KT \ln\left(2\sinh\frac{\beta\omega}{2}\right) \quad (25)$$

Once the Helmholtz free energy is obtained, the other statistical quantities are obtained in a straightforward manner. The mean energy is

$$U = -\frac{\partial}{\partial \beta} \ln Q_N = \frac{N\omega}{2} \coth \frac{\beta\omega}{2} \qquad (26)$$

The main statistical quantity, i.e., the entropy, is related to other quantities via

$$\frac{S}{K} = \beta^2 \left(\frac{\partial A}{\partial \beta}\right) = -N \ln\left(2 \sinh \frac{\beta\omega}{2}\right) + \frac{\beta N \omega}{2} \coth \frac{\beta\omega}{2} \qquad (27)$$

And the specific heat capacity at constant volume is obtained from

$$\frac{C}{K} = -\beta^2 \left(\frac{\partial U}{\partial \beta}\right) = \frac{\beta^2 \omega^2 N}{4} \left(\operatorname{csch} \frac{\beta\omega}{2}\right)^2 \qquad (28)$$

## 5- Information Entropy

The position space information entropies for the one-dimensional that can be calculated using $S_x = -\int |\Psi(x)|^2 \ln |\Psi(x)|^2 \, dx$. In general, explicit derivations of the information entropy are quite difficult. In particular the derivation of analytical expression for the $S_x$ is almost impossible. We represent the position information entropy densities respectively by $\rho_{S(x)} = |\Psi(x)|^2 \ln(|\Psi(x)|^2)$ [33-37]. $\rho_{S(x)}$ can be written as:

$$\rho_{S(x)} = \frac{m_0 e^{-\omega y^2} H_n(y\sqrt{\omega})^2 \ln\left(\frac{m_0 e^{-\omega y^2} H_n(y\sqrt{\omega})^2}{(1+\gamma x^2)^2}\right)}{(1+\gamma x^2)^2} \qquad (29)$$

The Fisher information entropies for the one-dimensional that can be calculated using $F_x = -\int |\Psi(x)|^2 \left[\frac{d}{dx} \ln |\Psi(x)|^2\right] dx$. In general, explicit derivations of the information entropy are quite difficult [38]. In particular the derivation of analytical expression for the $S_x$ is almost impossible. We represent the position information entropy densities respectively by $\rho_{F(x)} = |\Psi(x)|^2 \left[\frac{d}{dx} \ln |\Psi(x)|^2\right]$

Where $\rho_{F(x)}$ for can be written as:

$$\rho_{F(x)} = -\frac{4 m_0 e^{-\omega y^2} x\gamma \, H_n(y\sqrt{\omega})^2}{(1+\gamma x^2)^3} \qquad (30)$$

To understand treatments of these parameters, we have plotted some figures which illustrate these parameters such as what follows in below. At first case we plotted of $S_x$ considering $\gamma$ varying for $m_0 = 1$, $\omega = 1$. In Figs. (2) and (3), we depicted behavior of $F_x$ as function of $\gamma$ considering $m_0 = 1$, $\omega = 1$. The accending nature can be seen easily. In figure 4 and 5 we have plotted Shannon density versos x.

## 6- Some expectation values and the uncertainty principle

The asymmetry caused by the parameter $\gamma$ can be adequately quantified in terms of the average position of the particle in the box, calculated as $\langle x \rangle = \int_0^L \Psi^* x \Psi \, dx$ and average for $x^2$ is $\langle x^2 \rangle = \int_0^L \Psi^* x^2 \Psi \, dx$. From (19), and the average of the modified momentum is $\langle p \rangle = 0$ and average for $p^2$ is $\langle p^2 \rangle = -\hbar^2 \int_0^L \Psi^* \frac{d^2}{dx^2} \Psi \, dx$ [39, 40].
In the following, we have plotted uncertainty for x, p and the uncertainty principle. Figure 6 shows how uncertainty for x can change. Also treatments of uncertainty in momentum are shown in Fig-7 considering $m_0 = 1 \; and \; \omega = 1$ for different n in terms of $\gamma$. And in the last figure the assuming $m_0 = 1 \; and \; \omega = 1$, uncertainty principle is depicted as function of $\gamma$.

## 7- Conclusion

We studied a quasi-oscillator in Schrödinger equation within position-dependent mass formalism. In order to achieve our goal in article, we obtained wave function and energy spectra corresponding of the considered system, then, evaluated thermodynamic properties and information entropy as well as some expectation values and uncertainty principles. Also, we brought some numerical results as some graphs in some of section which illustrated what we derived and discussed in the article.

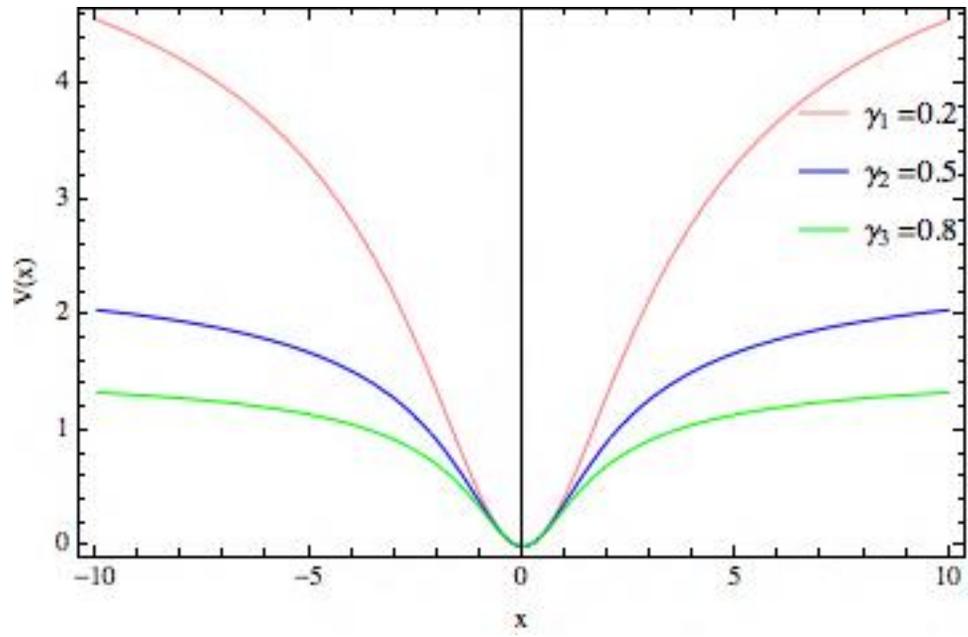

Fig-1 Behavior of $V(x)$ versus $x$ varying for $m_0 = 1$, $\omega = 1$

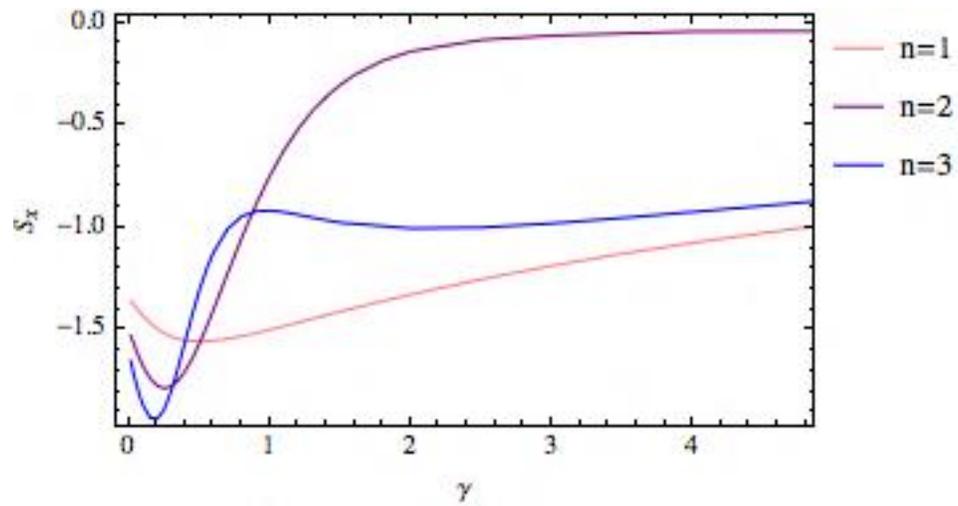

Fig-2 Behavior of $S_x$ versus $\gamma$ varying for $m_0 = 1$, $\omega = 1$

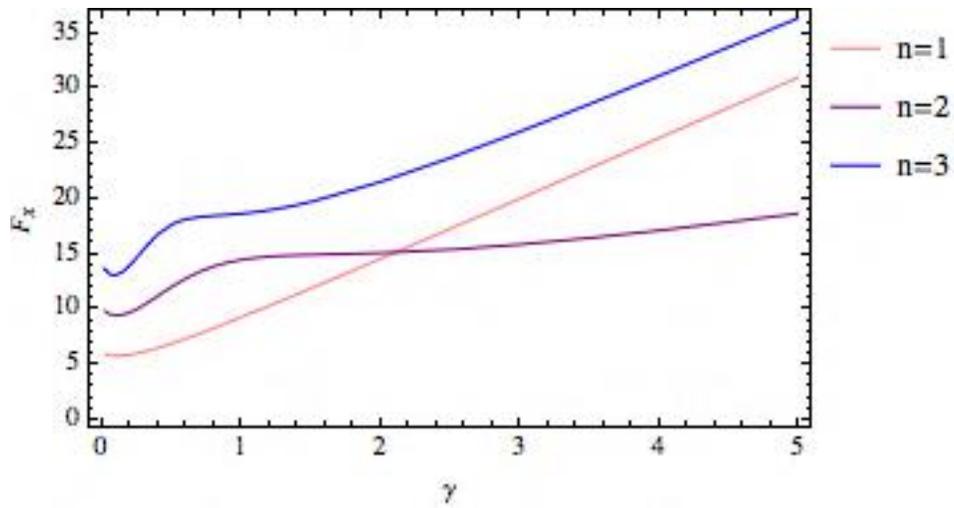

Fig-3 Behavior of $F_x$ versus $\gamma$ varying for $m_0 = 1$, $\omega = 1$

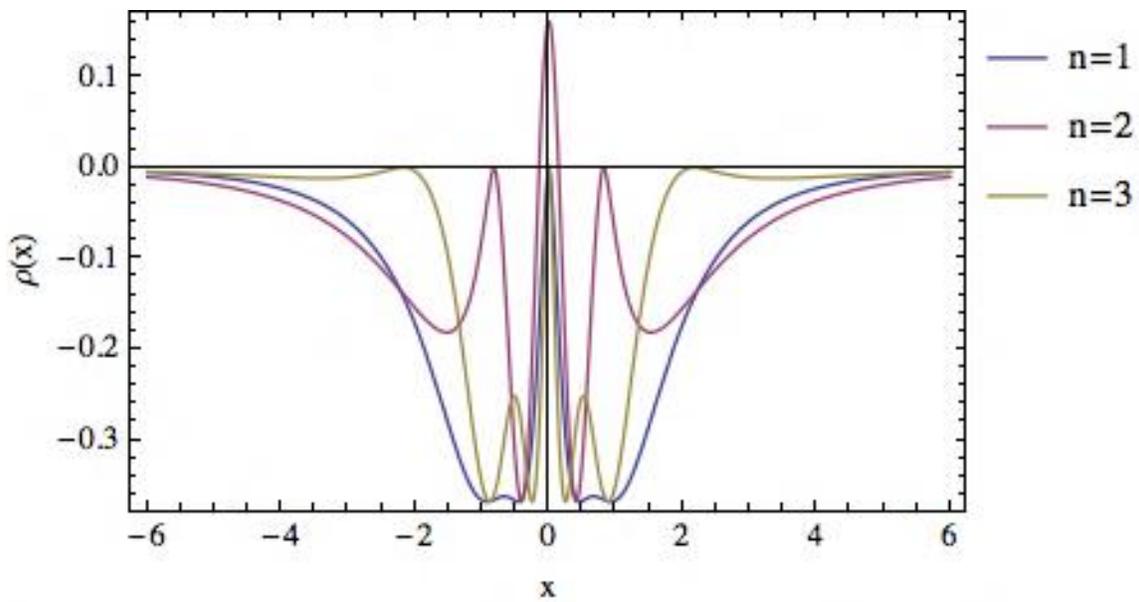

Fig-4 Behavior of $\rho(x)$ versus $x$ varying for $\gamma = 0.8$, $m_0 = 1$, $\omega = 1$

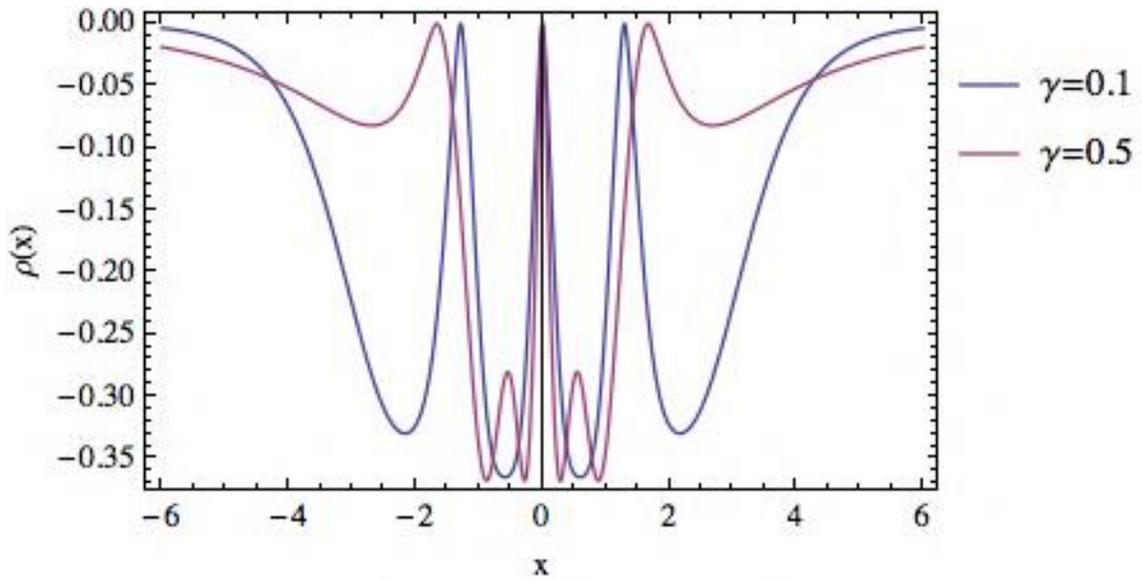

Fig-5 Behavior of $\rho(x)$ versus $x$ varying for $n = 3$, $m_0 = 1$, $\omega = 1$

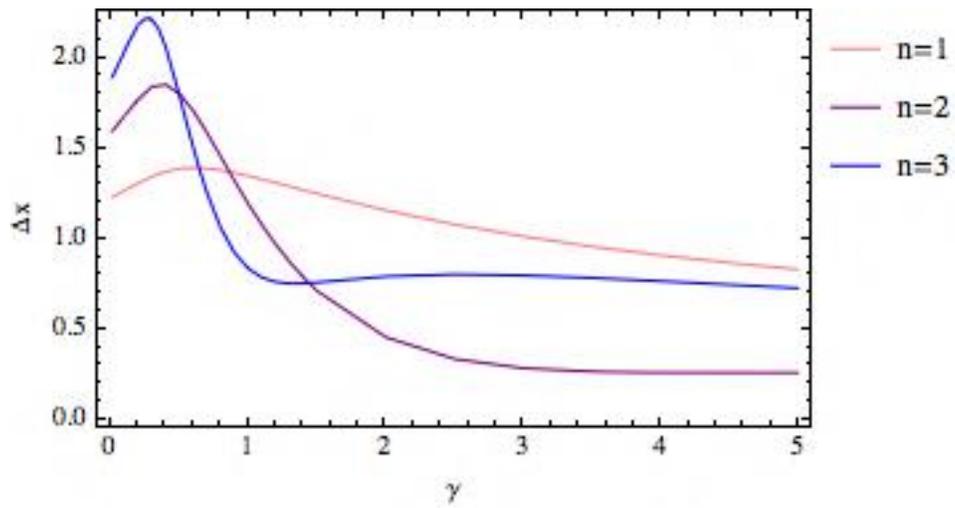

Fig-6 Uncertainty in position treatments varying for $m_0 = 1$, $\omega = 1$

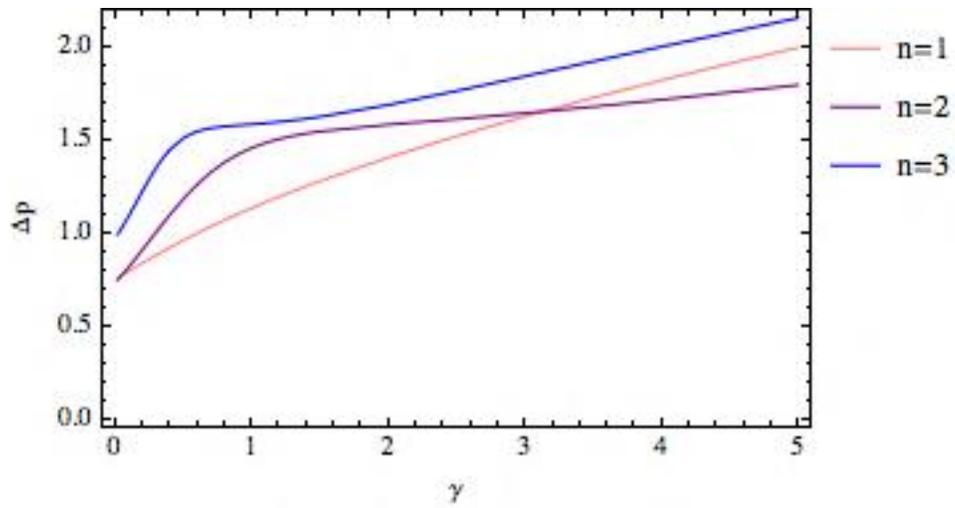

Fig-7 Uncertainty in momentum treatments varying for $m_0 = 1, \ \omega = 1$

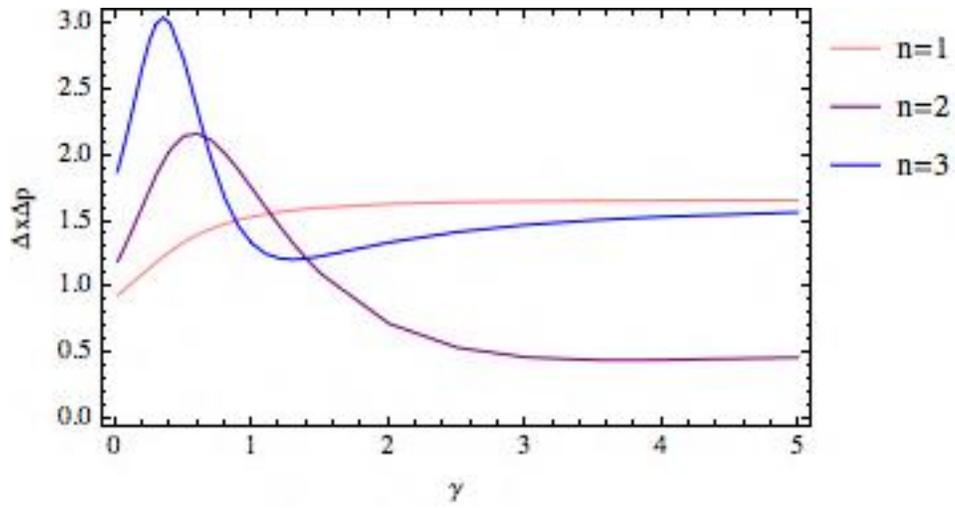

Fig-8 Uncertainty principle varying for $m_0 = 1, \ \omega = 1$

# References

[1] G. Bastard, *Wave Mechanics Applied to Semiconductor Heterostructure*, Les Editions de Physique, Les Ulis,France (1988).
[2] O. Von Roos and H. Mavromatis, Phys. Rev. B **31** (1985) 2294;
[3] Y.M. Li, H.M. Lu, O. Voskoboynikov, C.P. Lee, and S. M. Sze, Surf. Sci. **532** (2003) 811.
[4] A.J. Peter and K. Navaneethakrishnan, Physica E **40** (2008) 2747;
[5] M. Barranco, M. Pi, S.M. Gatica, E.S. Hernandez, and J. Navarro, Phys. Rev. B **56** (1997) 8997.
[6] F. Arias de Saavedra, J. Boronat, A. Polls, and A. Fabrocini, Phys. Rev. B **50** (1994) 4248.
[7] T.Q. Dai and Y.F. Cheng, Phys. Scr. **79** (2009) 015007;
[8] R. Renan, M.H. Pacheco, and C.A.S. Almeida, J. Phys. A: Math. Gen. **33** (2000) L509;
[9] L. Dekar, L. Chetouani, and T.F. Hammann, J. Math. Phys. **39** (1998) 2551;
[10] A.R. Plastino, A. Rigo, M. Casas, F. Garcias, and A. Plastino, Phys. Rev. A **60** (1999) 4318.
[11] V. Milanovic and Z. Ikovic, J. Phys. A: Math. Gen. **32** (1999) 7001.
[12] A.R. Plastino, A. Puente, M. Casas, F. Garcias, and A. Plastino, Rev. Mex. Fis. **46** (2000) 78.
[13] A. de Souza Dutra and C.A.S. Almeida, Phys. Lett. A **275** (2000) 25.
[14] A.R. Plastino, M. Casas, and A. Plastino, Phys. Lett. A **281** (2001) 297.
[15] B. Gön̈ul, O. Ozer, B. Gön̈ul, and F. Üzg̈un, Mod. Phys. Lett. A **17** (2002) 2453.
[16] C. Quesne and V. M. Tkachuk, J. Phys. A: Math. Gen. 37 (2004) 4267.
[17] L. Sierra and E. Lipparini, Euro. phys. Lett. 40 (1997) 667.
[18] F. S. A. Cavalcante, R. N. Costa Filho, J. Ribeiro Filho, C. A. S. de Almeida and C. N. Freire, Phys. Rev. B, 55 (1997) 1326.
[19] N. Bouchemla and L. Chetouani, Acta Phys. Pol. B 40 (2009) 2711
[20] C. Quesne, B. Bagchi, A. Banerjee and V. M. Tkachuk, Bulg. J. Phys. 33 (2006) 308.
[21] G. Ovand, J. Morales, J. J. Pena and G. Ares de Parga, Open Appl. Math. J 3 (2009) 29
[22] A. R. Plastino, M. Casas and A. Plastino, Phys. Lett. A 281 (2001) 297.
[23] Y. S. Cruz Cruz Trends in Mathematics 2013 (2013) 229.
[24] G. H. Sun, P. Dusan, C. N. Oscar and S. H. Dong, Chin. Phys. B 24, 10 (2015) 100303.
[25] B. J. Falaye, F. A. Serrano and S. H. Dong, Phys. Lett. A 380 (2016) 267–271
[26] N. Amir and Sh. Iqbal, Commun. Theor. Phys. **62** (2014) 790–794**.**
[27] F. D. Nobre and M. A. Rego-Monteiro, Braz J Phys 45(2015) 79.
[28] J. Yu and S. H. Dong, Phys. Lett. A 325 (2004) 194–198.
[29] J. Yu and S.H. Dong, G.H. Sun, Phys. Lett. A 322 (2004) 290.
[30] S. H. Dong and Z.Q. Ma, Int. J. Mod. Phys. E 11 (2) (2002) 155;
    L.Y. Wang, X.Y. Gu, Z. Q. Ma and S.H. Dong, Foun. Phys. Lett. 15 (2002) 569

[31] A. R. Nagalakshmi, B. A. Kagali, Physica Scripta Phys.Scr 77(2008) 015003
[32] R. K. Pathria, Statistical Mechanics, Pergamon press, Oxford, UK (1972)
[33] G. H. Sun, S. H. Dong and S. Naad, Ann. Phys. 525 (2013) 934.
[34] G. H. Sun, M. Avila Aoki and S. H. Dong, Chin. Phys. B 22 (2013) 050302.
[35] S. Dong, G. H. Sun,S. H. Dong and J. P. Draayer, Phys. Lett. A 378 (2014) 124.
[36] G. H. Sun, S.H. Dong, K. D. Launey, T. Dytrych and J. P. Draayer, Int. J., Quant.Chem. **115** (2015) 891–899.
[37] R. Valencia-Torres, G. H. Sun, S. H. Dong, Phys. Scr. **90** (2015) 035205.
[38] D. X. Macedo and I. Guedes, Physica A 434 (2015) 211–219.
[39] R. N. Costa Filho, M. P. Almeida, G. A. Farias, and J. S. Andrade Jr, Phys. Rev. A 84 (2011) 050102.
[40] M. A. Rego-Monteiro and F. D. Nobre , Phys. Rev. A 88 (2013) 032105.